\documentclass[10pt,aps,pra,reprint,longbibliography]{revtex4-1}
\pdfoutput=1

\usepackage[utf8]{inputenc}
\usepackage[T1]{fontenc}
\usepackage{graphicx}
\usepackage{color}

\usepackage{mathtools}
\usepackage{amssymb}
\usepackage{amsfonts}
\usepackage{dsfont}
\usepackage{esdiff}
\usepackage{braket}
\usepackage{hyperref}

\newcommand{\dd}{d}
\newcommand{\ii}{i}
\newcommand{\Trace}[1]{\text{tr}\left\{#1\right\}}     
\newcommand{\mat}[1]{\boldsymbol{#1}}                  

\begin{document}

\title{Optimal Qubit Control Using Single-Flux Quantum Pulses}
\author{Per J. Liebermann}
\email{per@lusi.uni-sb.de}
\author{Frank K. Wilhelm}
\affiliation{Theoretical Physics, Saarland University, Campus E2\,6, 66123 Saarbr\"ucken, Germany}

\begin{abstract}
  Single flux quantum pulses are a natural candidate for on-chip control of superconducting qubits. We show that they can drive high-fidelity single-qubit rotations---even in leaky transmon qubits---if the pulse sequence is suitably optimized. We achieve this objective by showing that, for these restricted all-digital pulses, genetic algorithms can be made to converge to arbitrarily low error, verified up to a reduction in gate error by 2 orders of magnitude compared to an evenly spaced pulse train. Timing jitter of the pulses is taken into account, exploring the robustness of our optimized sequence. This approach takes us one step further towards on-chip qubit controls.
\end{abstract}

\maketitle

\section{Introduction}
Rapid single flux quantum (RSFQ) technology has been and is originally pursued as an ultra-high-speed classical computing platform~\cite{Likharev1991,Likharev2012}. The ability to generate reproducible identical pulses at a high clock rate has been demonstrated in integrated circuits~\cite{Castellano2007}. Next to its original motivation of ultrafast digital circuits, this ability makes RSFQ technology a highly viable candidate for the on-chip generation of control pulses and readout for quantum computers based on Josephson devices~\cite{Orlando2002,Feldman2001,Ohki2007,Semenov2003,Fedorov2007,Fedorov2014}. The switching time lies in the picosecond range, leading to fast quantum gates~\cite{Ohki2005}, but the timing of the single-voltage pulses to control the devices is a major challenge~\cite{Gaj1997}. The integration capabilities of RSFQ technology are compatible with scaling quantum processors \cite{McDermott2014,Fowler2012}; for example, one can load pulse sequences into shift registers \cite{Mukhanov1993}. Then again, the present-day control scheme is based on room-temperature electronics, whose signals are transmitted as analog signals through filters into a cryostat, which creates large physical overhead.

Besides these engineering considerations, both techniques present distinct paradigms for control design. Room-temperature generators have high-amplitude resolution (currently, 12 bits are typical) at limited speed; thus, methods of analog pulse shaping can be applied~\cite{Glaser2015} to approximating the experiment with high accuracy. This approximation has been done in superconducting qubits, e.g., within the derivative removal by adiabatic gate (DRAG) and wah-wah method~\cite{Motzoi2009,Schutjens2013}, that suppress leakage into higher energy levels. These pulses can be readily calibrated using a protocol named Ad-HOC~\cite{Egger2014a}.

On the other hand, SFQ pulses only have a single bit of amplitude resolution: in a given time interval, there is either a pulse or not. The RSFQ sequence proposed in Ref.~\cite{McDermott2014} again reveals the challenge of balancing gate speed with suppression of leakage in transmon qubits similar to DRAG, demonstrating the need for advanced pulse design methods which cannot be solved by evenly spaced pulses alone.
While their amplitude resolution is minimal, an RSFQ sequence has a very short time constant of the pulse train---much shorter than the intrinsic frequencies of the qubit system---that can be expected to simulate analog control. To explore this matter, one needs to depart from the conventional optimal-control paradigms as reviewed in Ref.~\cite{Glaser2015}, which often involve gradients, and focus instead on digital control and algorithms that can optimize these fully discrete controls.

In this paper we show that optimal-control methods, based on genetic algorithms adapted to digital control, can be used to improve gate operations with trains of SFQ pulses. The single bit of amplitude resolution is encoded in a discretized time evolution, which also limits possible clock frequencies. We show that uneven and relatively densely populated bit-string pulse sequences can be used to suppress leakage as well as to drastically reduce gate times up to the speed limit. We also verify that optimized pulse sequences are robust against timing jitter, the degree of which depends on the clock integration.

\section{SFQ control}

\begin{figure}[]
  \centering
  \includegraphics[width=\columnwidth]{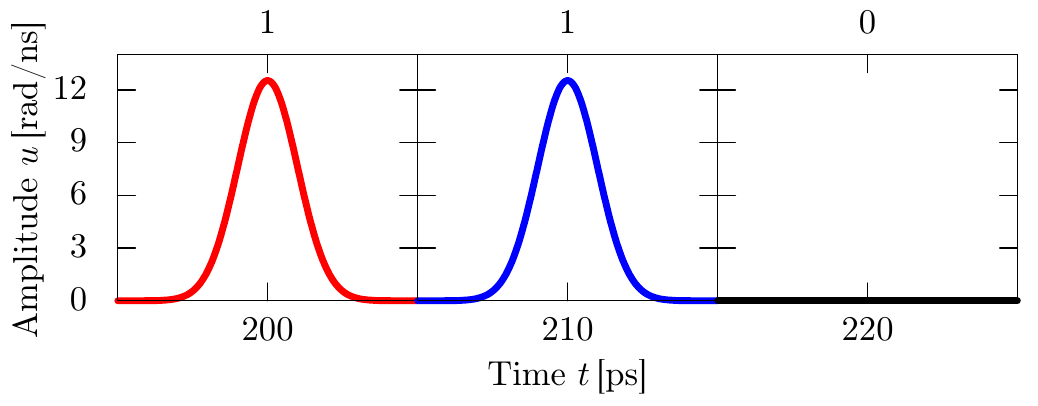}
  \caption{(Color online) Time slicing of the evolution: in each frame, the
    control amplitude $u(t)$ either has a Gaussian shape (1) or vanishes (0).
    Therefore, only two unitary operators $\mat{U}_i$ need to be calculated,
    each with a duration of $2t_c$ (here, 10\,ps). The gap between two
    consecutive pulses is always an integer multiple of $2t_c$.}
  \label{fig:cs}
\end{figure}

\begin{figure*}[]
  \centering
  \includegraphics[width=\textwidth]{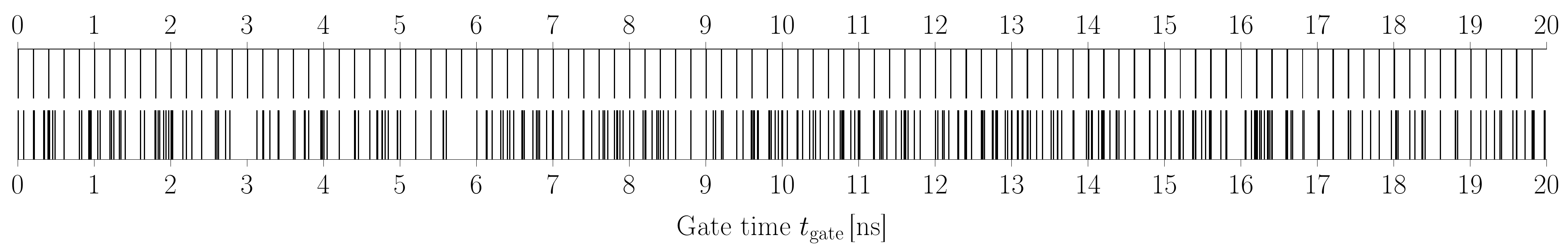}
  \caption{Bit-string representation of the pulse sequence before (top) and after (bottom) the optimization for a gate time of 20\,ns. A vertical black line indicates the time when a pulse is applied, and white space the duration of the free evolution.}
  \label{fig:bitstrings}
\end{figure*}

As a simple model, we use a qubit with a leakage level such as the transmon qubit~\cite{Koch2007,Girvin2011} that can be controlled with voltage pulses. Our single-qubit target gate is a rotation around the $y$~axis, i.e., a Pauli-$Y$ gate with an arbitrary global phase. The time-dependent system Hamiltonian in units of $\hbar$ reads
\begin{equation}
  \mat{H}(t) = \mat{H}_{0} + u(t)  \mat{H}_{1} \,,
  \label{eq:Hamiltonian}
\end{equation}
with drift and control terms
\begin{equation}
  \mat{H}_{0} =
  \begin{bmatrix}
    0 & 0 & 0 \\
    0 & \omega & 0 \\
    0 & 0 & 2\omega + \Delta
  \end{bmatrix}\,,\quad
  \mat{H}_{1} =
  \begin{bmatrix}
    0 & \frac{-\ii}{2} & 0 \\
    \frac{\ii}{2} & 0 & \frac{-\ii}{\sqrt{2} }\\
    0 & \frac{\ii}{\sqrt{2} } & 0
  \end{bmatrix}\,,
  \label{eq:Hcontrol}
\end{equation}
where $\omega$ is the qubit's angular frequency and $\Delta$ the anharmonicity of the third level. Gate operations are performed by changing the control amplitude $u(t)$. However, instead of finding a proper pulse shape with a duration of several nanoseconds, we apply a single-picosecond pulse repeatedly. This pulse is switched on and off as stored in a shift register and controlled by a clock. The pulse shape we use is a truncated Gaussian with standard deviation $\tau$ and total duration~$2t_c$
\begin{equation}
  u(t) = \frac{\delta\theta}{\sqrt{2\pi}\tau} e^{-t^2/2\tau^2} \qquad
  \int_{-t_c}^{t_c} u(t) dt \approx \delta\theta \,.
\end{equation}
The area $\delta\theta$ is approximated by the numerical integration of the pulse shape. It matches the rotation angle on the Bloch sphere for an infinitely narrow $\delta$~pulse. To flip the Bloch vector around the $y$~axis, we can thus set a lower bound of $n=\pi/\delta\theta$ pulses. Note that the three-level model outlined in Eq.~(\ref{eq:Hcontrol}) is sufficient to describe the qubit dynamics --- we perform {\it a posteriori} verifications with more levels and find no discernible difference.

We simulate our system by slicing the total gate time into $N$ time steps of length $2t_c$, i.e., the clock period. Choosing $\tau$ accordingly ensures that the applied voltage pulse shape $u(t)$ vanishes at the beginning and the end of a the time interval. We interpret the free evolution of the system in a time interval $[t_i, t_i+2t_c]$ as a pulse with zero amplitude. Figure~\ref{fig:cs} shows an example of two
consecutive pulses followed by an interval without an applied pulse. The time difference between two applied pulses is always a positive integer multiple of the pixel length $2t_c$. For each interval $[t_i, t_i+2t_c]$, the time evolution is captured in a unitary matrix $\mat{U}(t_i)\equiv \mat{U}(t_i+2t_c,t_i)$. Since we apply a single pulse only if it is necessary, each unitary is chosen out of a given database containing just two unitary operators, $\mat{U}_0$ and $\mat{U}_1$. For both our work and future practical applications, this can be done efficiently by storing all of the relevant system parameters in this database. Therefore, the pulse sequence can be represented as bit string, where zeros represent a free-evolution interval $\mat{U}_0$ and ones an applied voltage pulse $\mat{U}_1$; see Fig.~\ref{fig:bitstrings}. The total time evolution reads
\begin{equation}
  \mat{U}(t_\textrm{gate}) = \prod_{i=N-1}^0 \mat{U}(t_i) = \mat{U}(t_{N-1}) \cdots \mat{U}(t_0) \,.
\end{equation}
The database that the unitaries are chosen from consists of
\begin{align}
  \mat{U}_0 &= \exp\left( -2\ii t_c \mat{H}_{0} \right)
    = \mat{U}_0 (2t_c) \\
  \mat{U}_1 &= \mat{U}_0(t_c) \left[
    \mathcal{T} \exp\left( -\ii \int_{-t_c}^{t_c} \mat{H}(t) \dd t\right)
    \right] \mat{U}_0(-t_c) \,.
  \label{eq:database}
\end{align}
Any adjustment in the experiment can be captured in the database and all characterization of the pulses has to be done once in order to find these two database entries.

The target gate is a Pauli-$Y$ gate, where we allow an arbitrary phase for the leakage level, and the global phase is neglected when using an appropriate fidelity function [Eq.~(\ref{eq:phi})],
\begin{equation}
  \mat{U}_\mathrm{target} =
  \begin{bmatrix}
    0 & -1 & 0 \\
    1 & 0 & 0 \\
    0 & 0 & e^{\ii\varphi}
  \end{bmatrix}\,.
  \label{eq:target}
\end{equation}
We optimize our time evolution within the computational subspace of the qubit. The average fidelity function therefore reads~\cite{Rebentrost2009}
\begin{equation}
  \Phi = \frac{1}{4} \left|
    \Trace{\mat{U}^\dagger_\mathrm{target} \mat{P}_Q \mat{U}(t_\mathrm{gate}) \mat{P}_Q}
    \right|^2 \,,
  \label{eq:phi}
\end{equation}
with the projector onto the qubit subspace $\mat{P}_Q$.


Typical values for a transmon qubit~\cite{Koch2007} are for the qubit transition frequency $\omega/2\pi = 5$~GHz and its anharmonicity $\Delta/2\pi = -200$~MHz. The pulse area and duration are set to $\delta\theta = \pi/100$ and $2t_c=10$~ps, respectively, limiting the clock frequency to 100\,GHz. The gate time chosen is $t_g=20$~ns, leading to a total number of $N=2000$ pixels.

As a starting point, we use the sequence presented in Refs.~\cite{Bodenhausen1976,McDermott2014}, so we have exactly $n=100$ pulses in the beginning. After each pulse we wait $t_\mathrm{wait}=2\pi/\omega-2t_c$ for the qubit to complete a full precession before applying another pulse. This scheme increases the gate fidelities with an increasing number of applied pulses, while decreasing the area underneath the pulse shape with the same rate. It also leads to longer gate times, however, because every additional pulse increases the gate time by $2\pi/\omega$ due to the waiting time.

\section{Genetic algorithms}

\begin{figure}[t]
\centering
\includegraphics[width=\columnwidth]{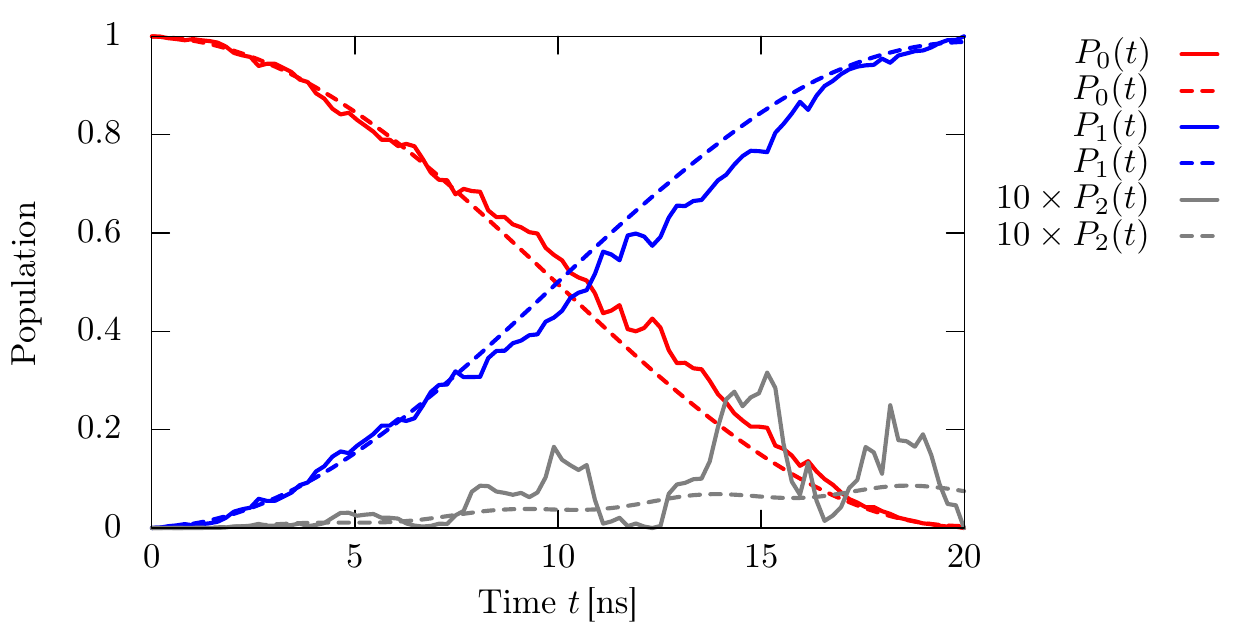} \\
\includegraphics[width=\columnwidth]{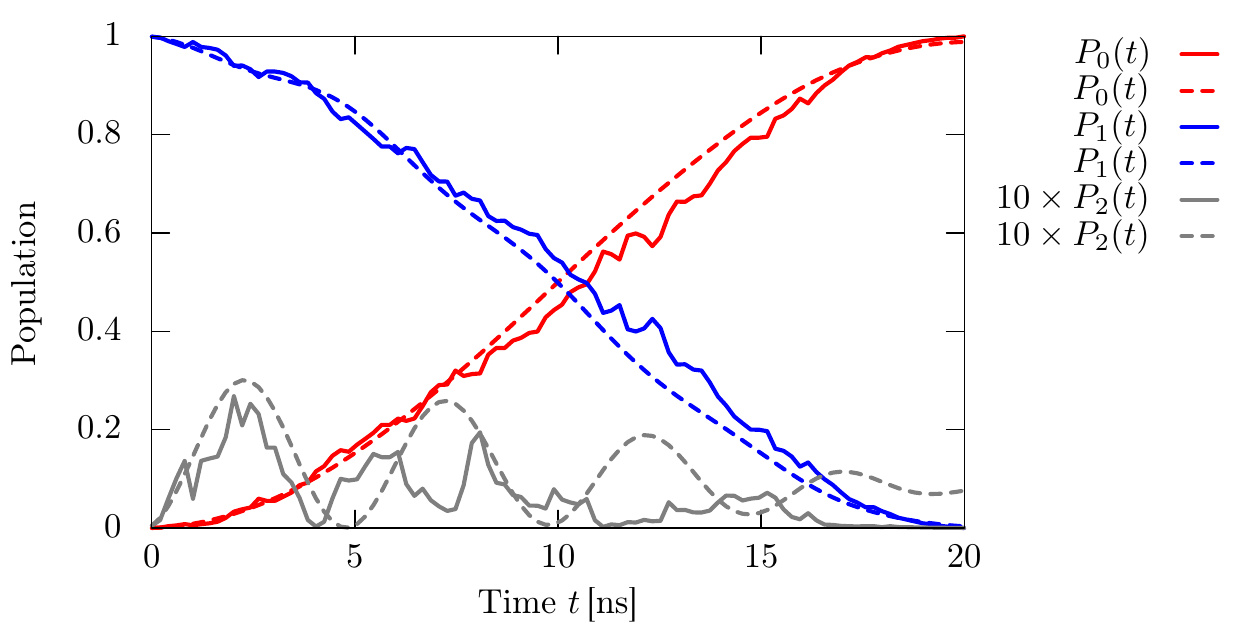}
\caption{Populations of a qutrit starting in the ground (top panel) and excited (bottom panel) state for the optimized sequence (solid lines) and the initial sequence (dashed lines). The population of the leakage level is enhanced by a factor of 10 for better legibility. The algorithm suppresses leakage to the ending gate time.}
  \label{fig:populations}
\end{figure}

As gradient-based algorithms are not straightforward here, we use a genetic algorithm~\cite{Whitley1994,Sutton1994} to optimize the pulse sequence. This versatile tool for global optimization is a natural candidate for the problem at hand, since our pulse sequence is already encoded in a binary string. Other genetic algorithms have been applied successfully in analog quantum optimal control, e.g., to optimize laser pulses to control molecules~\cite{Judson1992}.

Here, we search for a local minimum in the control landscape starting with the sequence described earlier~\cite{Bodenhausen1976,McDermott2014}, and stop as soon as we reach the target fidelity. Within the genetic algorithm framework, every solution for the variational parameters of the control problem is encoded in a genome. At each iteration, a selection of genomes will be merged by a crossover of genome pairs. The new and old genomes are mutated and the genomes with the best fitness make it to the next generation. Finding the right parameters for the genetic algorithm can be difficult, but it is common practice for most optimization problems to choose a high crossover and a low mutation probability~\cite{Hartmann2002}. The parameters of our optimization are shown in Table~\ref{tab:ga}.
\begin{table}[b]
 \begin{tabular}{l | r}
 \hline
 Population size & 70 \\
 \hline
 Mutation probability & 0.001 \\
 \hline
 Crossover probability & 0.9 \\
 \hline
 Number of genomes to select for mating & 64 \\
 \hline
 Maximum allowed iterations & 200\,000 \\
 \hline
 Target fitness & 0.9999 \\
 \hline
 Elitism & 1 \\
 \hline
 \end{tabular}
 \caption{Parameters used in the genetic algorithm. See Ref.~\cite{Whitley1994} for background.}
 \label{tab:ga}
\end{table}

\begin{figure}
\centering
\includegraphics[width=\columnwidth]{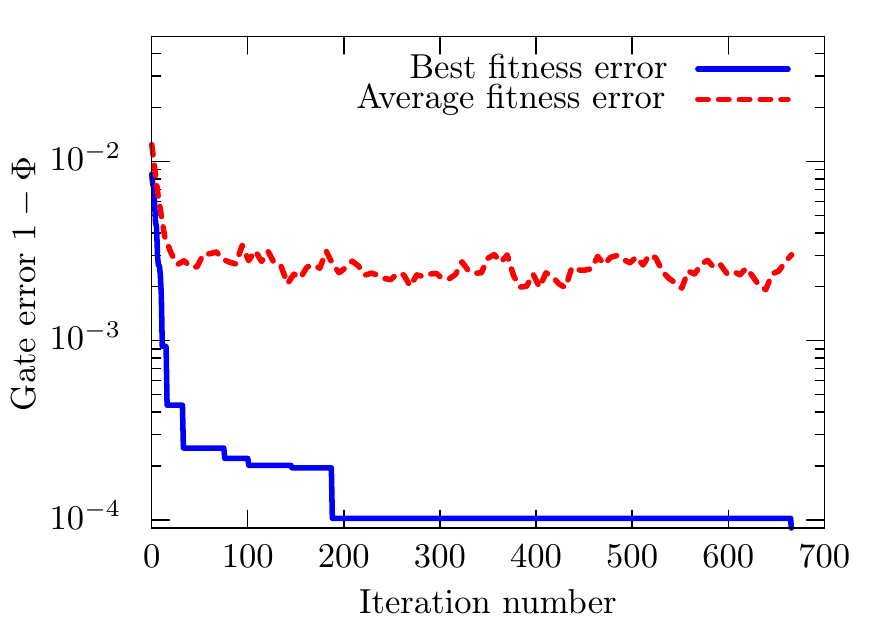}
\caption{Starting with a sequence shown in Ref.~\cite{Bodenhausen1976,McDermott2014},
 genetic algorithms are used for optimization. At each iteration, a new population of different pulse sequences is created from their parents, while the best solution of a generation with the lowest fidelity error is kept until either the algorithm finds a better one or a given threshold is exceeded.}
  \label{fig:error}
\end{figure}

Using that algorithm and setting our gate fidelity as a fitness measure we found the solutions for the sequence shown in Fig.~\ref{fig:bitstrings} and for the populations shown in Fig.~\ref{fig:populations}. The algorithm mainly corrects for leakage into the third energy level, which always leads to an increase in the number of pulses---in the solutions presented here, from $n=100$ to $n=301$. The time taken for a run is about 150\,s and the improvement is shown in Fig.~\ref{fig:error}. We encounter a broad variation over different runs indicating the presence of an abundance of local traps. This is consistent with the common observation that, in principle, trap-free control landscape~\cite{Rabitz2004} develops traps when the space of the available pulse shapes is strongly constrained.

\section{Quantum speed limit}

\begin{figure}
  \centering
  \includegraphics[width=\columnwidth]{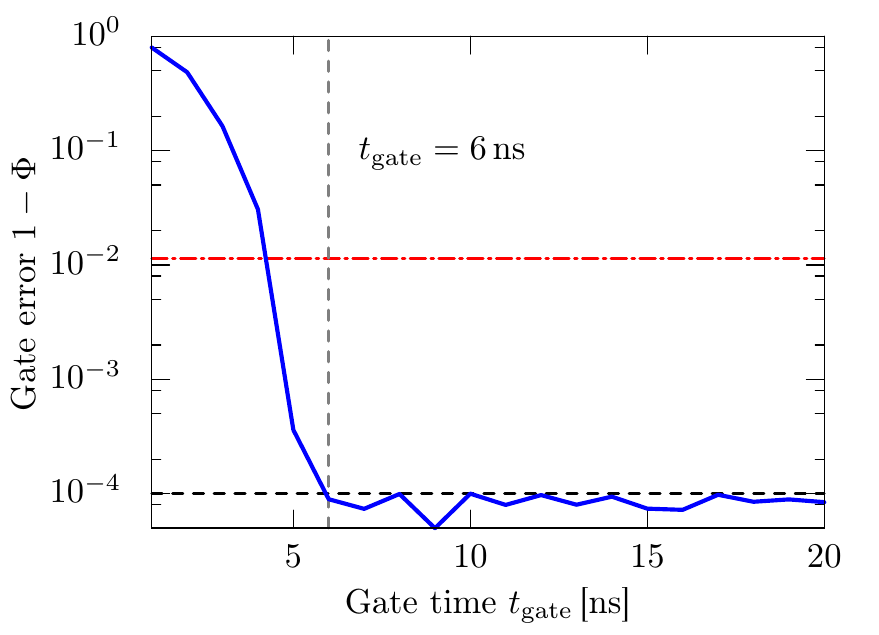}
  \caption{The gate errors of the optimizations for different gate times. The widths of the pulses have been kept constant; i.e., the number of pixels changes with the gate time linearly. The horizontal red dash-dotted line is the fidelity error of the initial sequence for 20~ns, and the horizontal black dashed line indicates the stopping condition of the algorithm when a fidelity $> 0.9999$ has been reached.}
  \label{fig:qsl}
\end{figure}

With the genetic algorithm at hand, we can search for shorter gate times $t_\textrm{gate}$ for the Pauli-$Y$ gate. We keep the widths of the pulses constant, which decreases the number of pixels with a decreasing gate time. The reachable fidelities are shown in Fig.~\ref{fig:qsl}. The shortest possible gate time we could find within the genetic algorithm is $t_\textrm{gate}=6$\,ns. Each optimization stops if a fidelity $>0.9999$ is reached or the maximum number of iterations (200\,000) is exceeded. We point out that, for short gate times, the evenly spaced pulse sequence is no longer a viable solution if we do not have any control of the pulse amplitude.

\section{Timing errors}

So far the clock has been assumed to be a perfect one. Here, we take inevitable timing errors into account, that lead to small pulse delays~\cite{Rylyakov1999} and, therefore, to deviations from the optimal fidelity. We simulate timing errors by multiplying every applied pulse $\mat{U}_1$ in the optimized sequence with a free-evolution operator of a time interval $\delta t$ from the right, and its adjoint from the left. Therefore, a positive $\delta t$ indicates a pulse which arrives with a delay, and a negative $\delta t$ indicates that the pulse arrives earlier:
\begin{equation}
  \mat{U}_1' = \mat{U}_0(-\delta t) \mat{U}_1 \mat{U}_0 (\delta t) \,.
  \label{eq:jitter}
\end{equation}

\begin{figure}
  \centering
  \includegraphics[width=\columnwidth]{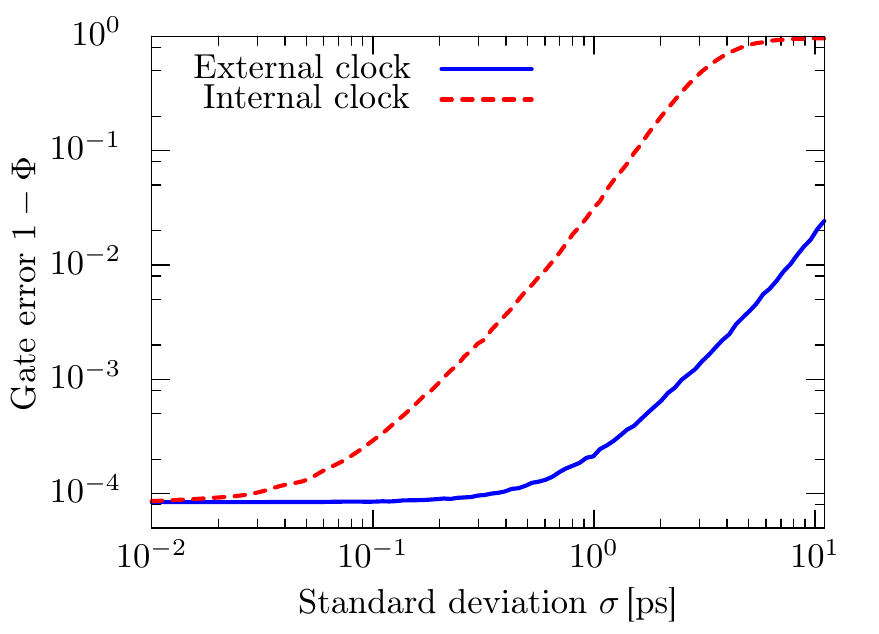}
  \caption{The gate error of the optimized
sequence for timing jitters with constant variance (external clock) and linear
growing variance (internal clock). The gate errors have been averaged over 1000
runs of the time evolution for each value of $\sigma$.}
  \label{fig:jitter}
\end{figure}

$\delta t$ is a normal distributed random time with standard deviation $\sigma$ for external jitter and $\sqrt{k} \sigma$ for internal jitter, where $k$ is the applied pulse number. For each value of $\sigma$, the fidelity of the time evolution has been averaged over 1000 runs for the optimized sequence. As can be obtained from Fig.~\ref{fig:jitter}, the external clocking scheme is more robust by an order of magnitude of the standard deviation. It is still within the target fidelity when the jitter time scale is 10\,\% of the pulse width, while, for an internal clock it is around 1\,\%. If the jitter time reaches the pulse duration, the gate error is still on the same scale as where we started our optimization. We therefore conclude that an external clock should be used in favor of an internal one for future devices.

\section{Conclusion}
We successfully develop and apply an optimal-control method for pulses with only a single bit of amplitude resolution. Finding the right binary string leads to minimization of the leakage error in the transmon system, and thus gate-control precision compatible with the requirements of fault-tolerant quantum computing. The results presented here show a fidelity improvement of several orders of magnitude over equal pulse-spacing sequences while being robust under external timing jitter. RSFQ shift registers are needed to perform the optimized sequence and are an essential part of on-chip SFQ-qubit control. This makes the underlying SFQ-pulse platform together with the single-bit optimal-control theory a possible and attractive candidate for an integrated control layer in a quantum processor.\\

\section*{acknowledgments}

We thank Daniel J. Egger and Oleg A. Mukhanov for the fruitful discussions.
Thiswork is supported by U.S. Army Research Office Grant No.~W911NF-15-1-0248.
This work is also supported by the EU through SCALEQIT.

\bibliography{literature}

\end{document}